%% file: arxiv.tex
\documentclass[ba,preprint]{imsart}

\RequirePackage{amsthm,amsmath,amsfonts,amssymb}
\RequirePackage[round]{natbib}
\RequirePackage[colorlinks,citecolor=blue,urlcolor=blue,backref=page,backref=page]{hyperref}
\usepackage{float}
\usepackage{tikz}
\usepackage{graphicx}

\pubyear{20XX}
\volume{TBA}
\issue{TBA}
\firstpage{1}
\lastpage{1}

\startlocaldefs
\theoremstyle{plain}

\theoremstyle{definition}

\theoremstyle{remark}

\endlocaldefs

\begin{document}

\begin{frontmatter}
\title{A strategy to avoid particle depletion in recursive Bayesian inference}
\runtitle{Avoiding particle depletion in recursive Bayesian inference}

\begin{aug}
\author[A]{\fnms{Henry}~\snm{Scharf}\ead[label=e1]{hscharf@arizona.edu}\orcid{0000-0002-9266-4191}}
\address[A]{Department of Mathematics,
University of Arizona\printead[presep={,\ }]{e1}}
\runauthor{H. Scharf}
\end{aug}

\begin{abstract}
Recursive Bayesian inference, in which posterior beliefs are updated in light of accumulating data, is a tool for implementing Bayesian models in applications with streaming and/or very large data sets. As the posterior of one iteration becomes the prior for the next, beliefs are updated sequentially instead of all-at-once. Thus, recursive inference is relevant for both streaming data and settings where data too numerous to be analyzed together can be partitioned into manageable pieces. In practice, posteriors are characterized by samples obtained using, e.g., acceptance/rejection sampling in which draws from the posterior of one iteration are used as proposals for the next. While simple to implement, such filtering approaches suffer from particle depletion, degrading each sample's ability to represent its target posterior. As a remedy, we investigate generating proposals from a smoothed version of the preceding sample's empirical distribution. The method retains computationally valuable properties of similar methods, but without particle depletion, and we demonstrate its accuracy in simulation. We apply the method in simulation to both a simple, logistic regression model as well as a hierarchical model originally developed for classifying forest vegetation in New Mexico using satellite imagery.
\end{abstract}



\end{frontmatter}


\section{Introduction}\label{sec:introduction}
When both the number of unknown parameters in a statistical model and number of observations available to estimate those parameters are ``large'', the computational cost of model fitting can pose a substantial, sometimes insurmountable, burden to statistical inference. Likelihood-based approaches such as Bayesian and maximum likelihood estimation typically require multiple evaluations of the likelihood while exploring the high-dimensional parameter space, and when the number of observations is large, each evaluation can require large amounts of both processing resources and memory. The present work is motivated by an application of a species distribution model \cite[e.g.,][]{guisan_predicting_2005, GELF2006} in which both landscape characteristics (e.g., slope, elevation) and satellite imagery inform the estimated probabilities that a site is dominated by one of $K$ species of vegetation \citep{ASSA2015, scharf_2024}. The spatio-temporal and spectral extent of the application leads to tens of millions of observations from the available satellite imagery, which are used to learn hundreds of unknown parameters. A full implementation of the statistical model using all available observations is not possible due to constraints on the amount of random access memory required to compute the log-likelihood. One way to address this computational bottleneck is to adopt a recursive approach in which satellite imagery is incorporated sequentially in batches, without ever requiring all observations to be available at once.

\subsection{Recursive Bayesian inference}

It is helpful to establish notation that will aid in discussing the overall framework of recursive Bayesian inference. Let $\mathcal{P}^J_n$ be a partition of the index set $\{1, \dots, n\}$ into $J$ subsets of respective sizes $n_1, \dots, n_J$ such that $\sum_{j=1}^J n_j = n$ with a natural ordering suggested by the index $j = 1, \dots, J$. Let $\boldsymbol{y}$ denote the complete set of $n$ observations, $\{y_1, \dots, y_n\}$, and let $\boldsymbol{y}_{j}$ denote observations corresponding to the $j$th subset in $\mathcal{P}^J_n$ such that $\cup_{j = 1}^J \boldsymbol{y}_j = \boldsymbol{y}$. Further, let $\boldsymbol{y}_{1:j} = \cup_{i = 1}^j \boldsymbol{y}_i$ denote the sequence of nested sets for $j = 1, \dots, J$, each of size $\sum_{i = 1}^j n_i$. In addition, let $[\boldsymbol\theta]$ denote the probability distribution for the, possible multivariate, random variable $\boldsymbol\theta$. When it is necessary to avoid confusion, $[\boldsymbol\theta = \boldsymbol\theta^*]$ will be used to indicate the probability density for the random variable $\boldsymbol\theta$ evaluated at $\boldsymbol\theta^*$.

A recursive Bayesian approach to estimating the posterior distribution of parameters $\boldsymbol\theta$ given observations $\boldsymbol{y}$ supposes that some knowledge about the ``transient'' posterior density, $[\boldsymbol\theta | \boldsymbol{y}_{1:j-1}]$ is available and can be updated in light of $\boldsymbol{y}_j$ to yield knowledge about $[\boldsymbol\theta | \boldsymbol{y}_{1:j}]$. Updates proceed recursively until the ultimate posterior $[\boldsymbol\theta | \boldsymbol{y}]$ is known. In practice, transient posteriors are not available in closed-form and are instead represented using samples drawn from their distribution using Monte Carlo (MC) techniques.

In the Prior-Proposal-Recursive Bayesian (PP-RB) inference paradigm \citep{hooten_making_2021}, proposals from the joint transient posterior $[\boldsymbol\theta | \boldsymbol{y}_{1:{j - 1}}]$ are used in a Metropolis-Hastings (MH) algorithm for the $j$th of $J$ stages to obtain draws from $[\boldsymbol\theta | \boldsymbol{y}_{1:j}]$. The procedure makes use of a convenient simplification of the acceptance probability that arises when proposals come from the transient posterior. Let $[\boldsymbol\theta^* | \boldsymbol\theta^{t-1}]$ be the proposal density for a MH algorithm targeting $[\boldsymbol\theta | \boldsymbol{y}_{1:j}] \propto [\boldsymbol{y}_j, \boldsymbol\theta | \boldsymbol{y}_{1:j-1}]$ in which new values are accepted with probability $\alpha\left(\boldsymbol\theta^*, \boldsymbol\theta^{t-1}\right) = \min(1, r^t_j)$ with 
\begin{align}
r^t_j &= 
  \frac{[\boldsymbol{y}_j, \boldsymbol\theta = \boldsymbol\theta^*|\boldsymbol{y}_{1:j-1}]}
    {[\boldsymbol{y}_j, \boldsymbol\theta = \boldsymbol\theta^{t-1}|\boldsymbol{y}_{1:j-1}]}
  \frac{[\boldsymbol\theta^{t-1} | \boldsymbol\theta^*]}
    {[\boldsymbol\theta^* | \boldsymbol\theta^{t-1}]} \\
&= \frac{[\boldsymbol{y}_j | \boldsymbol\theta = \boldsymbol\theta^*, \boldsymbol{y}_{1:j-1}]
      [\boldsymbol\theta = \boldsymbol\theta^*|\boldsymbol{y}_{1:j-1}]}
    {[\boldsymbol{y}_j | \boldsymbol\theta = \boldsymbol\theta^{t-1}, \boldsymbol{y}_{1:j-1}]
      [\boldsymbol\theta = \boldsymbol\theta^{t-1}|\boldsymbol{y}_{1:j-1}]}
  \frac{[\boldsymbol\theta^{t-1} | \boldsymbol\theta^*]}
    {[\boldsymbol\theta^* | \boldsymbol\theta^{t-1}]}.
\end{align}
For $[\boldsymbol\theta^* | \boldsymbol\theta^{t-1}] = [\boldsymbol\theta = \boldsymbol\theta^*|\boldsymbol{y}_{1:j-1}]$, the MH ratio simplifies to
\begin{align}
r^t_j= \frac{[\boldsymbol{y}_j | \boldsymbol\theta = \boldsymbol\theta^*, \boldsymbol{y}_{1:j-1}]}
    {[\boldsymbol{y}_j | \boldsymbol\theta = \boldsymbol\theta^{t-1}, \boldsymbol{y}_{1:j-1}]}. \label{eqn:mh_ratio}
\end{align}
For models in which the observations $\boldsymbol{y}_j$ are conditionally independent given $\boldsymbol\theta$ \citep[CIHMs,][]{kass1989approximate, gelfandghosh2013}, the acceptance ratio simplifies to 
\begin{align}
r^t_j = \frac{[\boldsymbol{y}_j | \boldsymbol\theta = \boldsymbol\theta^*]}
  {[\boldsymbol{y}_j | \boldsymbol\theta = \boldsymbol\theta^{t-1}]},
\end{align}
highlighting a desirable feature of the approach that only observations $\boldsymbol{y}_j$ are directly involved during the $j$th stage. By carrying out all $J$ stages, it is possible to obtain exact samples from the posterior distribution $[\boldsymbol\theta | \boldsymbol{y}]$, yielding Bayesian inference on the parameters of interest.

In the PP-RB approach, proposals from the previous transient posterior distribution, $[\boldsymbol\theta | \boldsymbol{y}_{1:j-1}]$, are typically obtained by sampling from the empirical distribution of previous stage's sample. That is, for a sample of size $M$, $\{\boldsymbol\theta^i|\boldsymbol{y}_{1:j-1}\}_{i=1}^M$, drawn during stage $j-1$, the multinomial random variable $\boldsymbol\theta^*|\{\boldsymbol\theta^i|\boldsymbol{y}_{1:j-1}\}_{i=1}^M~\sim~\mathrm{Cat}\left(\{\boldsymbol\theta^i|\boldsymbol{y}_{1:j-1}\}_{i=1}^M\right)$ has the required marginal distribution, $[\boldsymbol\theta^*] = [\boldsymbol\theta = \boldsymbol\theta^*|\boldsymbol{y}_{1:j-1}]$. Thus, a recursive Bayesian inference algorithm can be implemented by sequentially updating realized samples from the previous transient posterior based on the appropriate MH ratio. 

The recursive approach was initially motivated by the goal of combining inference in a meta-analysis of several previous studies \citep{lunn_fully_2013}, but is also useful for reducing the memory demands of analyses involving very large $n$ because the $j$th stage only requires working with $n_j$ observations when the observations are conditionally independent.

\subsection{Particle depletion}

A challenge arising from the multinomial sampling approach in PP-RB is that the realized values in the samples from each successive transient posterior are necessarily a subset of the realized values in the preceding sample, and thus the number of unique values in the sequence of samples is monotonically non-increasing. As information contained in new observations is included, transient posteriors shift and contract, leading to low acceptance rates for many previous-stage realizations, and a rapid drop in the number of unique proposals. Those proposals that are retained across each stage have increasing multiplicity, assuming $M$ remains constant. Ultimately, posterior realizations collapse to a single value with multiplicity $M$.


The process through which the sequence of samples converge to a single repeated value is an example of the well known phenomenon of particle ``depletion'' (or ``degeneracy'') in the particle filtering literature, and it is the motivation behind a large body of research \cite[e.g.,][]{andrieu_particle_2010}. Especially when the support of a random variable is known to be continuous, particle depletion is problematic because it exacerbates MC error and degrades a sample's ability to adequately represent its associated target distribution resulting in excessive statistical uncertainty that worsens as the number of stages increases.

I propose a generalization of the PP-RB approach, called smoothed PP-RB (SPP-RB), in which MH proposals are drawn from a distribution over continuous support closely approximating the transient posterior. The approximation of the transient posterior is precise enough to allow the same cancellation in the PP-RB MH ratios, and also ensures that all proposed values are unique, avoiding particle depletion. The method is computationally inexpensive, especially relative to the cost of evaluating the transient likelihood $[\boldsymbol{y}_{j}|\boldsymbol\theta]$, resulting in a generalized algorithm with computing and memory demands on par with the original PP-RB. In simulation, the new method dramatically reduces the discrepancy between target posteriors and those obtained by traditional PP-RB using multinomial sampling.


A key benefit of the SPP-RB approach is that it naturally admits a procedure for block MH updates at each stage of the recursive algorithm unavailable with PP-RB. In general, PP-RB requires joint proposals from the transient posterior distribution, such that the entire proposal vector is either rejected or accepted at each iteration. For moderate- to high-dimensional parameter spaces, acceptance rates can drop rapidly with the dimensionality, and without any mechanism to tune the MH algorithms, particle depletion can become problematic after only a few stages. SPP-RB naturally extends to MH algorithms with blocking structures design to improve acceptance rates resulting in more efficient exploration of the parameter space.

%

\section{Perturbations to avoid particle depletion}\label{sec:perturb}
A natural inclination in response to particle depletion is to try and break the increasing multiplicity arising from the multinomial proposals by introducing small, random perturbations. For parameters with continuous probability densities, adding a small amount of extra variation will have a correspondingly small effect on the resulting probability density and thus the perturbed random variables will have a distribution approximately equal to the transient posterior. With some care, such an approach can indeed be shown to yield unique values with approximately the desired distribution.

Let $\{\boldsymbol\theta\}^M = \{\boldsymbol\theta^1, \dots, \boldsymbol\theta^M\}$ be a sample of size $M$ with each $\boldsymbol\theta^i \sim [\boldsymbol\theta], i = 1, \dots, M$. Define a new random variable, $\boldsymbol\theta^*$, with distribution, $[\boldsymbol\theta^*] \approx [\boldsymbol\theta]$ via the following conditional probability distributions:
\begin{align}
\boldsymbol{\tilde\theta}|\{\boldsymbol\theta\}^M &\sim \mathrm{Cat}\left(\{\boldsymbol\theta\}^M\right) \label{eqn:sample}\\
\boldsymbol\theta^*|\boldsymbol{\tilde\theta}, \{\boldsymbol\theta\}^M
  &\sim \left[\boldsymbol\theta^* | \boldsymbol{\tilde\theta}, \{\boldsymbol\theta\}^M\right]. \label{eqn:perturb}
\end{align}
Under this paradigm, $\boldsymbol\theta^*$ arises by first selecting a member of the sample, $\{\boldsymbol\theta\}^M$, completely at random then perturbing the realization according to $\left[\boldsymbol\theta^* | \boldsymbol{\tilde\theta}, \{\boldsymbol\theta\}^M\right]$. For example, when $\left[\boldsymbol\theta^* | \boldsymbol{\tilde\theta}, \{\boldsymbol\theta\}^M\right] = \mathrm{N}(\boldsymbol\theta^*| \boldsymbol{\tilde\theta}, \sigma^2\mathbf{I})$, where $\mathrm{N}(\boldsymbol{x}| \boldsymbol\mu, \boldsymbol\Sigma)$ denotes the multivariate normal density with mean $\boldsymbol\mu$ and covariance matrix $\boldsymbol\Sigma$ evaluated at $\boldsymbol{x}$, perturbing transient posterior samples is accomplished by adding spherical Gaussian white noise. 

For the anticipated application in recursive Bayesian inference, desirable properties of $\boldsymbol\theta^*$ are (i) a high probability of drawing unique values under repeated sampling, even when $\{\boldsymbol\theta\}^M$ may contain several duplicated values, and (ii) $[\boldsymbol\theta^*] \approx [\boldsymbol\theta]$. Property (i) is assured with probability 1 if $\boldsymbol\theta^*| \boldsymbol{\tilde\theta}, \{\boldsymbol\theta\}^M$ has continuous support. One approach for ensuring property (ii) is attempting to match the moments of $\boldsymbol\theta^*$ as closely as possible with those of $\boldsymbol\theta$. In particular, it is possible to specify a practical procedure that ensures the first two moments of $\boldsymbol\theta^*$ match those of $\boldsymbol\theta$ asymptotically as $M\rightarrow\infty$.

\subsection{Sampling from a kernel density estimate}
An intuitively appealing candidate for the perturbation distribution, \eqref{eqn:perturb}, is the Gaussian distribution centered on $\boldsymbol{\tilde\theta}$ with covariance matrix $\mathbf{H}_\theta$ denoted by $\left[\boldsymbol\theta^*|\boldsymbol{\tilde\theta}, \{\boldsymbol\theta\}^M\right] = \mathrm{N}(\boldsymbol\theta^*| \boldsymbol{\tilde\theta}, \mathbf{H}_\theta)$. Such a specification is equivalent, marginally, to drawing $\boldsymbol\theta^*|\{\boldsymbol\theta\}^M$ from a kernel density estimate (KDE) based on $\{\boldsymbol\theta\}^M$ with Gaussian kernel and bandwidth matrix determined by $\mathbf{H}_\theta$. The KDE is a mixture distribution with density 
\begin{align}
[\boldsymbol\theta^* | \{\boldsymbol\theta\}^M] = M^{-1}\sum_{i = 1}^M \mathrm{N}\!\left(\boldsymbol\theta^*| \boldsymbol{\theta}^i, \mathbf{H}_\theta\right).
\end{align}
Gaussian KDEs are among the most widely used tools for density estimation, particularly for multivariate densities \citep{chacon_multivariate_2018}, and are thus an attractive possibility for approximating the transient posteriors. As the two step procedure given by \eqref{eqn:sample}--\eqref{eqn:perturb} makes clear, it is straightforward and computationally inexpensive to sample from such a mixture for many families of perturbation distributions such as the multivariate Gaussian.

For this KDE-based strategy, $\mathrm{E}(\boldsymbol\theta^*) = M^{-1} \sum_{i=1}^M \mathrm{E}(\boldsymbol\theta^i) = \mathrm{E}(\boldsymbol\theta)$. Although $\mathrm{Var}(\boldsymbol\theta^*) = \mathrm{Var}(\boldsymbol\theta) + \mathbf{H}_\theta > \mathrm{Var}(\boldsymbol\theta)$, in practice most KDE bandwidth selection methods have the property that $\lim_{M\rightarrow\infty} \mathbf{H}_\theta = \mathbf{0}$, so the second moments align asymptotically. The limiting case of a zero covariance matrix for a finite sample size is equivalent to the multinomial resampling used in PP-RB \citep{hooten_making_2021}.

Unfortunately, unconstrained multivariate bandwidth selection for more than a few variables at a time is known to be a very difficult problem for which no optimal, computationally efficient methods currently exist \citep{chacon_multivariate_2018}. Constraining the space of bandwidth matrices to, for example, diagonal matrices, yields methods for high dimensional bandwidth estimators, but does not perform well in practice (see Section~\ref{sec:simulation-study}). 

\subsection{Assuming joint Gaussianity}\label{sec:joint_gauss}

An alternative approach is to abandon the non-parametric perspective underpinning KDEs and instead assume that each transient posterior is well represented by a Gaussian distribution. Rather than introduce small perturbations to the empirical distribution, let $[\boldsymbol\theta^*|\{\boldsymbol\theta\}^M] = \mathrm{N}(\boldsymbol\theta^*| \boldsymbol{\bar\theta}, \mathbf{S}_\theta)$, where $\boldsymbol{\bar\theta}$ denotes the sample mean of $\{\boldsymbol\theta\}^M$. Specifying $\mathbf{S}_\theta$ to be an unbiased estimator of $\mathrm{Var}(\boldsymbol\theta)$ ensures alignment of the first two moments. 

There are two reasons for optimism about this approach. First, unlike KDE bandwidth matrices, variance-covariance estimators can be computed efficiently for high dimensional spaces. Second, the Bernstein-von Mises theorem ensures that, under mild regularity conditions, transient posteriors asymptotically approach Gaussianity as sample sizes increase \citep[e.g.,][]{hartigan_asymptotic_1983}. 

\subsection{Regularized KDE}\label{sec:reg_KDE}

A compromise between the non-parametric perspective underpinning multinomial resampling and the parametric assumption of joint Gaussianity, is to regularize (or ``shrink'') the values obtained in \eqref{eqn:sample} toward a common Gaussian distribution through the following perturbation distribution:
\begin{align}
\left[\boldsymbol\theta^*| \boldsymbol{\tilde\theta}, \{\boldsymbol\theta\}^M\right] &= \mathrm{N}\!\left(\boldsymbol\Lambda \boldsymbol{\tilde\theta} + 
  (\mathbf{I} - \boldsymbol\Lambda)\boldsymbol{\bar\theta}, 
  \mathbf{S}_\theta - 
    \boldsymbol\Lambda\mathbf{S}_\theta\boldsymbol\Lambda^\prime\right), \label{eqn:reg_KDE}
\end{align}
where $\mathbf{S}_\theta$ is an asymptotically unbiased estimate of the variance-covariance matrix for $\boldsymbol\theta$ based on $\{\boldsymbol\theta\}^M$, such as the sample covariance matrix, and $\boldsymbol\Lambda$ is a square matrix. The marginal distribution of $\boldsymbol\theta^* | \{\boldsymbol\theta\}^M$ is the mixture
\begin{align}
M^{-1}\sum_{i = 1}^M \mathrm{N}\!\left(\boldsymbol\theta^*|
  \boldsymbol\Lambda \boldsymbol\theta^i + 
    (\mathbf{I} - \boldsymbol\Lambda)\boldsymbol{\bar\theta}, 
  \mathbf{S}_\theta - \boldsymbol\Lambda\mathbf{S}_\theta\boldsymbol\Lambda^\prime\right). 
  \label{eqn:reg_mix}
\end{align}

It is straightforward to verify that the marginal expectation $\mathrm{E}(\boldsymbol\theta^*) = \mathrm{E}(\boldsymbol\theta)$. In addition, application of the law of total variation can be used to show that $\lim_{M\rightarrow\infty}\mathrm{Var}(\boldsymbol\theta^*) = \mathrm{Var}(\boldsymbol\theta)$ (see Appendix). Thus, for sufficiently large sample sizes, drawing $\boldsymbol\theta^*$ according to \eqref{eqn:reg_KDE} ensures correct alignment of the first two moments.

A useful, simple form for $\boldsymbol\Lambda$ is $\boldsymbol\Lambda = \lambda\mathbf{I}$, where $\lambda$ takes on a value in the closed unit interval, $[0, 1]$. The conditional mean of $\boldsymbol\theta^*$ for this setting is a convex combination of $\boldsymbol{\tilde\theta}$ and the sample mean $\boldsymbol{\bar\theta}$. The variable $\lambda$ can be viewed as a regularization or tuning parameter, where $\lambda = 0$ corresponds to an expected value shrunk all the way to $\boldsymbol{\bar\theta}$ and is equivalent to the assumption of joint Gaussianity for transient posteriors in Section~\ref{sec:joint_gauss}, and $\lambda = 1$ corresponds to no shrinkage whatsoever, equivalent to multinomial resampling. 

\begin{figure}[h]
\includegraphics[width = \textwidth]{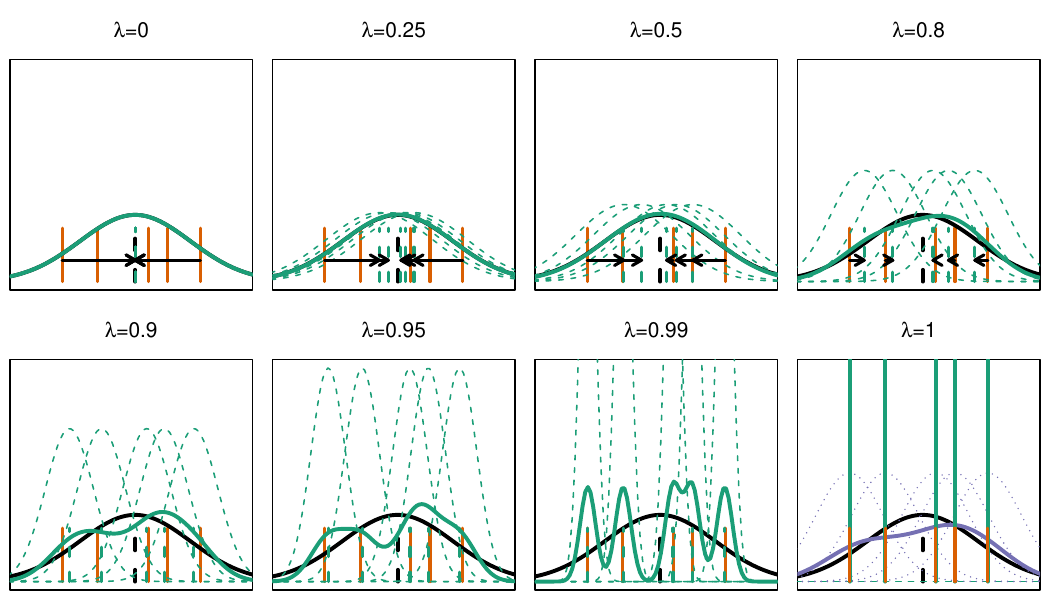}
\caption{Schematic representation of the proposed regularized KDE for varying values of the regularization parameter, $\lambda$. Orange vertical lines represent realizations from an unknown distribution. The solid green line represents the density estimate for the particular value of $\lambda$, which is a mixture of the densities depicted with green dashed lines. In the bottom right figure, the purple solid line shows an un-regularized KDE, with components shown as purple, dashed lines.}
\label{fig:schematic_reg_KDE}
\end{figure}

Figure~\ref{fig:schematic_reg_KDE} shows a schematic for the proposed regularized KDE in one dimension. As $\lambda$ varies from 0 (fully Gaussian) to 1 (empirical distribution of $\{\theta\}^M$), the realized values of $\theta^*$ (green, dashed vertical lines) move from $\bar{\theta}$ (black, dashed vertical line) toward realizations of $\tilde\theta$ drawn from $\mathrm{Cat}\left(\{\theta\}^M\right)$ (orange, solid vertical lines). In addition, their associated components in the resulting regularized KDE (green, dashed curves) become more peaked, eventually approaching point masses. An un-regularized KDE based on $\{\theta\}^M$ using a common bandwidth selection convention is shown in purple in the bottom left plot (purple, dashed curves show the components).

\section{Smoothed prior-proposal recursive Bayes}

Equipped with an approximate procedure for obtaining samples from the continuous distribution underlying a finite sample, PP-RB can be extended to use values during stage $j$ proposed according to \eqref{eqn:sample} and \eqref{eqn:reg_KDE}. Because MH proposals at each stage are drawn from a distribution representing a smoothed version of an empirical distribution, the extended version of PP-RB is called ``smoothed prior-proposal recursive Bayesian'' inference, or SPP-RB. In what follows, $\boldsymbol\Lambda = \lambda \mathbf{I}$ is assumed to be a scalar multiple of the identity matrix, with $\lambda \in [0, 1]$, with $\lambda = 0$ equivalent to jointly Gaussian proposals, and $\lambda = 1$ corresponding to multinomial samples, or PP-RB. The SPP-RB approach outperformed both original PP-RB and un-regularized KDEs with diagonal bandwidth structure in the simulation studies explored in Sections~\ref{sec:logistic} and \ref{sec:spatial_multi}.

\subsection{Blocking in SPP-RB}

Conventional PP-RB implements MH algorithms at each stage in which all variables are jointly proposed from the transient posterior and all accepted or rejected. For high dimensional models, non-negligible values of the acceptance ratio can be highly concentrated in the parameter space, leading to acceptance rates that decrease as the number of dimensions grows. Low acceptance rates leads to inefficient MH algorithms and can further worsen the problem of particle depletion. In conventional MCMC applications, MH algorithms are often implemented within a Gibbs sampling scheme so that subsets, or blocks, of variables can be considered separately \citep[e.g.,][]{gelman_bayesian_2014}. Blocking can increase acceptance rates and effective sample sizes, and is therefore a sampling technique that could enable recursive Bayesian inference algorithms in applications with high-dimensional models.

Let $\mathcal{P}_P^B$ be a partition of the index set $\{1, \dots, P\}$ into $B$ subsets that describes a blocking structure on the full vector of model variables, $\boldsymbol\theta$, where $P$ corresponds to the dimensionality of $\boldsymbol\theta$ and $B$ the number of blocks. Define $\boldsymbol\theta_b$ to be the sub-vector of $\boldsymbol\theta$ corresponding to the $b$th block, $b = 1, \dots, B$ and $\boldsymbol\theta_{-b}$ to be the complementary sub-vector of all elements not the $b$th block. Standard MH within Gibbs provides samples from the joint distribution of $\boldsymbol\theta$ by using MH algorithms to sample from the ``full-conditional'' distribution of a block of variables conditioned on all complementary variables outside the block, $\left[\boldsymbol\theta_b | \boldsymbol\theta_{-b}, \boldsymbol{y}\right]$ \citep[e.g.,][]{gelman_bayesian_2014}.

In the context of recursive Bayesian inference, the target for the $b$th block of a Gibbs update during each iteration is 
\begin{align*}
[\boldsymbol\theta_b | \boldsymbol\theta_{-b}, \boldsymbol{y}_{1:j}] \propto 
  [\boldsymbol{y}_j, \boldsymbol\theta_b | \boldsymbol\theta_{-b}, \boldsymbol{y}_{1:j-1}]
= [\boldsymbol{y}_{j} | \boldsymbol\theta, \boldsymbol{y}_{1:j-1}]
  [\boldsymbol\theta_b | \boldsymbol\theta_{-b}, \boldsymbol{y}_{1:j-1}].
\end{align*}
For a MH algorithm with proposal distribution $[\boldsymbol\theta^*_b | \boldsymbol\theta^*_{-b} = \boldsymbol\theta^{t-1}_{-b}, \boldsymbol{y}_{1:j-1}]$, where $\boldsymbol\theta^{t-1}_{-b}$ are the most recent values of $\boldsymbol\theta_{-b}$ in the broader Gibbs algorithm, cancellations in the MH ratio yield $r^t_j = \frac{[\boldsymbol{y}_j | \boldsymbol\theta^*, \boldsymbol{y}_{1:j-1}]}{[\boldsymbol{y}_j | \boldsymbol\theta^{t-1}, \boldsymbol{y}_{1:j-1}]}$ as in \eqref{eqn:mh_ratio}, which further simplifies to $r^t_j = \frac{[\boldsymbol{y}_{j} | \boldsymbol\theta^*]}{[\boldsymbol{y}_{j} | \boldsymbol\theta^{t-1}]}$ when the observations, $\boldsymbol{y}$, are conditionally independent. 

Proposing from the transient full conditional $[\boldsymbol\theta_b | \boldsymbol\theta_{-b}, \boldsymbol{y}_{1:j-1}]$ could in principle proceed via multinomial sampling from $\{\boldsymbol\theta : \boldsymbol\theta_{-b} = \boldsymbol\theta^{t-1}_{-b}\}^M$, the subset of transient posterior samples for which the out-of-block component matches the most recent values in the MH algorithm. However, this subset is generally very small in practice, perhaps even of size 1, limiting exploration of the parameter space. The effect worsens in the presence of particle depletion and for increasing number of blocks, $B$, effectively rendering blocking ineffective with multinomial sampling.

For SPP-RB, proposals can be made from the conditional distribution of \eqref{eqn:reg_mix}, which can be shown to be a weighted mixture of Gaussian distributions. Let $\boldsymbol\mu^i = \boldsymbol\Lambda \boldsymbol\theta^i + (\mathbf{I} - \boldsymbol\Lambda)\boldsymbol{\bar\theta}$ and $\boldsymbol\Sigma = \mathbf{S}_\theta - \boldsymbol\Lambda \mathbf{S}_\theta \boldsymbol\Lambda$. Then  
\begin{align}
\left[\boldsymbol\theta^*_b | \boldsymbol\theta^*_{-b} = \boldsymbol\theta_{-b}^{t-1}, \{\boldsymbol\theta\}^M\right] = 
    \sum_{i = 1}^M w_i \mathrm{N}\left(\boldsymbol\theta^*_b| \boldsymbol\mu^i_{b|-b},
  \boldsymbol\Sigma_{b|-b} \right), \label{eqn:cond_reg} \\
  \boldsymbol\mu^i_{b|-b} = \boldsymbol\mu^i_b + 
  \boldsymbol\Sigma_{b, -b}\boldsymbol\Sigma^{-1}_{-b, -b}
    (\boldsymbol\theta_{-b}^{t-1} - \boldsymbol\mu^i_{-b}), 
  \quad \boldsymbol\Sigma_{b|-b} = \boldsymbol\Sigma_{b, b} - 
    \boldsymbol\Sigma_{b, -b}\boldsymbol\Sigma^{-1}_{-b, -b}\boldsymbol\Sigma_{-b, b},
\end{align}
where subscripts on $\boldsymbol\mu^i$ and $\boldsymbol\Sigma$ indicate the subvectors and submatrices associated with the indices, and the weights are
\begin{align}
w_i \propto \mathrm{N}\left(\boldsymbol\theta^*_{-b}| \boldsymbol\mu^i_{-b},
  \boldsymbol\Sigma_{-b,-b}\right). \label{eqn:weights}
\end{align}
In practice, sampling from \ref{eqn:cond_reg} can be implemented like most mixture distribution samplers by first computing the weights, $w_i$, then selecting a mixture component index, $i^*$ with probability proportional to $w_i$, and finally drawing a realization from the corresponding component. 

Blocking is generally more computationally intensive per iteration than single, joint updates, because it requires $MB$ multivariate Gaussian density evaluations for the weights, and $B$ likelihood calculations. However, the computational expense is often worth the gains in efficiency because fewer total iterations are necessary to obtain sufficient samples from each transient posterior.

It is noteworthy that the case of $\lambda = 0$ requires less computation, since for this situation the conditional density $\left[\boldsymbol\theta^*_b | \boldsymbol\theta_{-b}^{t-1}, \{\boldsymbol\theta\}^M\right]$ is also multivariate normal. Thus, if normality can be assumed for each transient posterior, the SPP-RB algorithm is especially efficient. 




\subsection{Diagnostic checks}\label{sec:diagnostics}
A clear concern in implementing any recursive approach for inference is ensuring that the final stage accurately reflects the holistic learning one would have if the entire data set were available for analysis all-at-once. The simulation studies in Section~\ref{sec:simulation-study} permit direct comparison with the natural reference of an all-at-once analysis, and so it is possible to verify that the recursive approach ultimately arrives at approximately the correct destination. However, in practice, such a comparison will not generally be possible.

Because the proposed method for recursive Bayesian inference is theoretically valid for any partition, $\mathcal{P}_n^J$, and sequencing, one way to check for discrepancy between samples obtained via a recursive algorithm and the true target distributions is to carry out the recursive procedure with multiple distinct partitions and compare the results. While \textit{transient} posteriors for different partition designs will not generally target the same distributions and therefore cannot be meaningfully compared, the ultimate posteriors must coincide. Thus, if samples drawn from the final stages of recursive Bayesian algorithms based on different partition designs appear to arise from different distributions, it may be concluded that the final samples from at least one of the recursive algorithms do not faithfully represent the true target posterior. In contrast, when all partition designs yield indistinguishable samples, it is reasonable to conclude that the samples arise from the target distribution. Figure~\ref{fig:logistic_diagnostics} shows an example of how multiple recursive fits can be used to generate a diagnostic plot for assessing convergence to the target distribution. When different partition sequences yield substantially different posteriors, remedial measures such as increasing the stage-wise sample size, $M$, or adjusting $\lambda$ may improve performance.

\section{Simulation studies}\label{sec:simulation-study}

\subsection{Logistic regression}\label{sec:logistic}
The performance of SPP-RB was investigated for the simple setting of logistic regression where it could be directly compared with sample posteriors obtained using an all-at-once algorithm that would typically be inaccessible for real applications. An intercept and five synthetic predictor variables, $\boldsymbol{x}_i$, were simulated independently from a multivariate normal distribution with identity covariance matrix
for $i = 1, \dots, n$ observations. Binary realizations were then simulated according to a logistic regression model such that
\begin{align}
[\boldsymbol\beta] &= \mathrm{N}(\boldsymbol\beta| \mathbf{0}, \sigma_\beta^2 \mathbf{I}), \\
[y_i | \boldsymbol\beta] &= 
  \mathrm{Bern}\left(\mathrm{logit}^{-1}(\boldsymbol{x}_i^\prime \boldsymbol\beta)\right), 
  \quad i = 1, \dots, n
\end{align}
where the multivariate normal density for $[\boldsymbol\beta]$ represents prior beliefs about the regression coefficients, $\boldsymbol\beta$. A sample of size $n = 380$ was partitioned into a first batch of size 50 and 11 subsequent batches of size 30 for sequential analysis. Posterior samples were obtained using SPP-RB for $\lambda = 0, 0.25, 0.5, 0.75, 0.9, 0.95, 0.99, 1$ as well as for a recursive Bayesian approach where proposals were drawn from a KDE based on transient posterior samples using a diagonal bandwidth matrix with entries chosen marginally according to 5 commonly used bandwidth selectors. For each cumulative collection of observations, $\boldsymbol{y}_{1:j}, j = 1, \dots, 12$, samples from the transient posterior were also obtained using an all-at-once MCMC algorithm. Finally, 16 separate stage 1 samples of size $M = 10,000$ were generated after 2,000 warm-up iterations using random initializations to the no U-turn sampler of \cite{hoffman2014no} implemented in Stan \citep{carpenter2017stan} via the R package \texttt{brms} \citep{brms} based on the same simulated data. Recursive Bayes algorithms were implemented for all 16 sets of samples to help quantify variation across different stage 1 realizations. Performance of sequential algorithms was judged based on how well transient posteriors agreed with those obtained from the all-at-once approach. 

\begin{figure}[h]
\includegraphics[width=\textwidth]{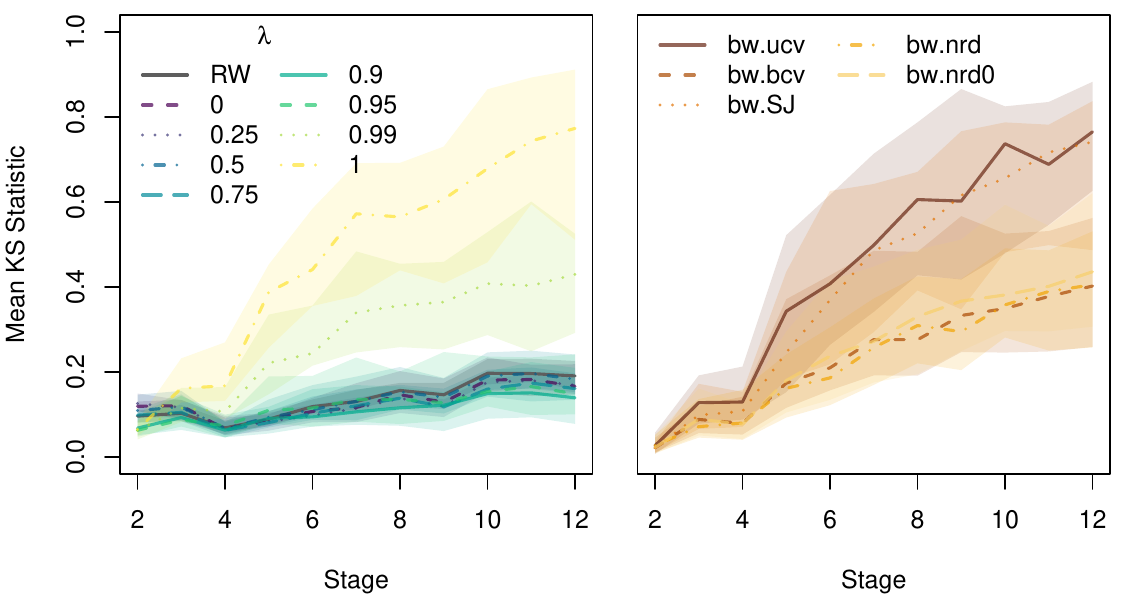}
\caption{Median of the mean KS statistics computed for each of 16 stage 1 samples (dark lines) and pointwise 2.5\% and 97.5\% quantiles.}
\label{fig:mean_KS}
\end{figure}

To summarize the discrepancy between the samples drawn from each transient posterior distribution using regularized SPP-RB and those drawn using the all-at-once approach, Kolmogorov--Smirnov (KS) statistics were computed marginally for each element of $\boldsymbol\beta$ and then averaged to yield a one-number summary for each posterior sample. Mean KS statistics were computed for all 16 repeated fits. Figure~\ref{fig:mean_KS} shows the median value of the mean KS statistics across the 16 fits (dark lines), along with pointwise 2.5\% and 97.5\% quantiles for each stage (transparent polygons). Large values indicate KS statistics associated with stronger evidence that transient posterior samples from the recursive approach differed from the all-at-once approach and hence poorer performance. 

The SPP-RB approach performed well for a range of $\lambda$ values from 0 to 0.9 (Figure~\ref{fig:mean_KS}, left). KDE approaches with diagonal bandwidth matrices did not perform as well as SPP-RB in general, although a few methods for selecting bandwidths (nrd, nrd0, and bcv) outperformed other alternatives within the category of marginal KDE (Figure~\ref{fig:mean_KS}, right). QQ-plots comparing marginal quantiles for recursive to the all-at-once approaches for a single replicate are included in the Appendix (Figures~\ref{fig:qq_spprb} and \ref{fig:qq_kde}) and support the broad conclusions drawn from Figure~\ref{fig:mean_KS}.

\begin{figure}[h]
\includegraphics[width=\textwidth]{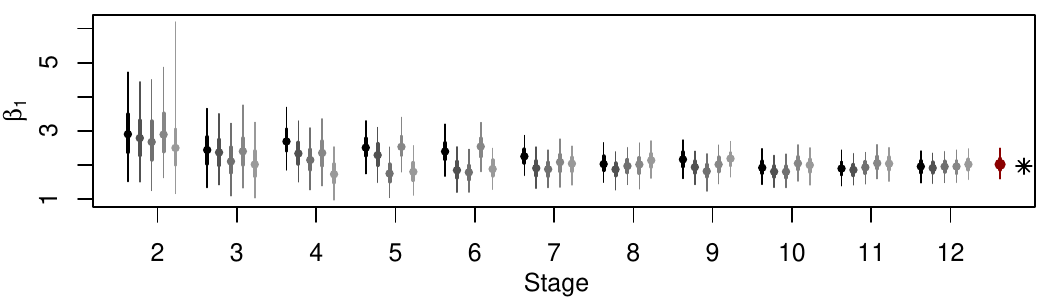}
\caption{Diagnostic plot showing variation in transient posteriors depending on partition sequence (gray box plots), and convergence to the correct ultimate posterior distribution (red box plot). The asterisk shows the true value of the regression coefficient used to simulate the data.}
\label{fig:logistic_diagnostics}
\end{figure}

As discussed in Section~\ref{sec:diagnostics}, one approach to assessing the validity of the ultimate posterior samples when an all-at-once baseline is not available is to carry out the recursive Bayesian approach using multiple partition schemes. To implement this diagnostic tool, 5 separate partition sequences with equal partition sizes were created and used to implement the SPP-RB approach via the 5 different resulting update pathways. Figure~\ref{fig:logistic_diagnostics} shows a summary of the marginal transient posteriors for one particular regression coefficient for the case of $\lambda = 0$, with shades of gray corresponding to the 5 unique pathways. Although sequences of transient posteriors can differ substantially in their trajectory depending on the paritculars of any one partition (e.g., stages 5 and 6), Figure~\ref{fig:logistic_diagnostics} shows that when a recursive algorithm is working properly, the ultimate posteriors for all partition sequencess coincide with each other. This reconciliation is visible in the figure as close agreement among marginal posteriors at stage 12, and also with the all-at-once baseline (dark red).

To minimize MC error in the target posterior, recursive inference algorithms rely on the existence of a sufficiently large sample of values to adequately represent each transient posterior. However, increasing the number of samples obtained during each stage also increases the computational demand of implementation. By comparing the discrepancy of ultimate posteriors based on distinct partition sequences, it is possible to determine whether MC error arising from an insufficient number of samples (i.e., too small a value for $M$) has been adequately controlled.

As the dimensionality of a target posterior grows, more samples, $M$, from each transient posterior are expected to be required to control MC errors do not accumulate. To investigate the impact of increasing model complexity on the success of the SPP-RB approach and requirements for $M$, the logistic regression example was expanded to include a growing number of regression coefficients, $P$. Samples from the ultimate posterior distributions were obtained using several different values for $M$, with 5 unique partitions implemented for each unique combination of $P$ and $M$, all for $\lambda = 0$. Pairwise KS statistics were calculated for each parameter across all 10 possible pairs of fits. For each pair of fits, the KS statistics for all parameters were averaged, yielding 10 values quantifying the agreement between each pair of ultimate posteriors.

\begin{figure}[h]
\includegraphics[width=\textwidth]{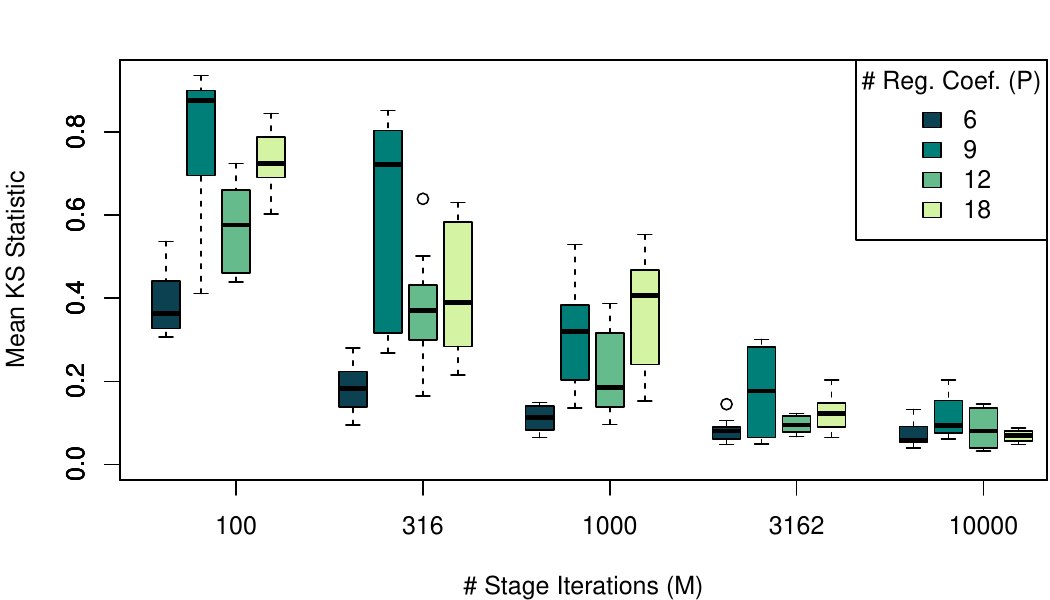}
\caption{Mean KS statistics across all model parameters. Mean KS statistics were computed for each of 10 possible pairs of ultimate posterior samples obtained from five unique partition sequences. Each boxplot corresponds to a particular combination of number of regression coefficients, $P$, and iterations per stage, $M$. Lower mean KS statistics indicate greater agreement among ultimate posteriors, and therefore suggest lower levels of MC error.}
\label{fig:KS_stat_logistic}
\end{figure}

Figure~\ref{fig:KS_stat_logistic} shows boxplots of mean KS statistics across all 10 pairwise combinations. For $P = 6$ predictors, as few as 1,000 iterations per stage leads to very close agreement between the final posteriors obtained using SPP-RB and those obtained using an all-at-once approach. However, as $P$ increases, more iterations are required.

\subsection{Species distribution model}\label{sec:spatial_multi}
In anticipation of applying the SPP-RB approach to the motivating application of species distribution modeling, a second simulation study involving a more complex hierarchical model involving a latent multinomial logistic regression component was also investigated. The model is designed to help predict the pattern of, and understand effects related to, forest vegetation in a system dominated by $K$ species. The true dominant vegetation species is only observed at a small number of sites due the expense associated with gathering such data. However, indirect observations of dominant species cover are available via satellite imagery. Images are recorded as the proportion of incident sunlight light reflected off the surface of the earth at a given site. Reflectances are measured at a discrete set of wavelengths throughout the year, with revisit intervals for each site dependent on the trajectory of the observing satellite relative to the site.

Sites are expected to reflect different amounts of light depending on the wavelength, day of the year, and dominant species associated with the site. Each species is associated with a unique profile of reflectances across wavelengths that also vary seasonally. For instance, deciduous tree species can be expected to reflect more light at visibly green wavelengths during the summer than during the winter. While similar species may have similar reflectance profiles, subtle differences in the timing of senescence (changing of leaf color and dropping of leaves) and magnitudes of reflectance across spectra can help distinguish dominant species type when many reflectance measurements are available at each site.

Logit-transformed reflectances associated with each pixel were modeled as conditionally independent Gaussian random variables with means given by a function of wavelength, day of the year, and primary vegetation cover type across the pixel. The conditional distribution of $r_{ij}$, the logit-transformed measurement of reflectance at site $j$ corresponding to wavelength $w_{ij}$ on date $d_{ij}$, was defined to be
\begin{align}
\left[r_{ij}|z_j = k, \boldsymbol\gamma_k, \sigma^2\right] &= 
  \mathrm{N}(r_{ij} | \boldsymbol{g}_{ij}^\prime\boldsymbol\gamma_{k}, \sigma^2), 
    \quad j \in \{1, \dots, J \}, i \in \{1, \dots, N_j\}. \label{eqn:reflectance}
\end{align}
Each $y_j \in \{1, \dots, K\}$ is a categorical random variable representing the dominant vegetation type associated with each of the $J$ sites. The vector $\boldsymbol{g}_{ij} = \boldsymbol{g}(w_{ij}, d_{ij})$ is the evaluation of a set of $L$ flexible two-dimensional basis functions, $\boldsymbol{g}(\cdot, \cdot)$, at a particular wavelength, $w_{ij}$ and date $d_{ij}$, and $\boldsymbol\gamma_{k}$ is an associated vector of regression coefficients indexed by the dominant species type $k$. Thus, the inner product $\boldsymbol{g}(\cdot, \cdot)^\prime\boldsymbol\gamma_{k}$ defines $K$ unique surfaces as functions of wavelength and time, one for each species. Species share basis functions, reflecting a shared space of possible reflectance surfaces, but have different sets of coefficients, $\boldsymbol\gamma_k$, allowing each species to have a reflectance surface that captures its own seasonal growing patterns. The variance parameter, $\sigma^2$, corresponds to residual, independent variation associated with each measurement attributable to observation error.

A multinomial logistic regression model links the dominant species at each site to landscape characteristics that explain variation in species' distributions and account for residual spatial autocorrelation. Let $\boldsymbol{x}_j^\prime$ denote a row-vector of $P$ predictor values corresponding to landscape characteristics at site $j$, and let $\boldsymbol\beta_k$ denote a vector of species-specific regression coefficients for each covariate. 
The probability, $\Pr(z_j = k) = p_{jk}$, of observing cover type $k$ at site $j$ is defined via the following two relationships
\begin{align}
\log(\eta_{jk}) &=
  \boldsymbol{x}^\prime_j \boldsymbol\beta_k
  ,
  \quad k \in \{1, \dots, K\} \label{eqn:eta} \\
p_{jk} &= \frac{\eta_{jk}}{1 + \sum_{k = 1}^{K - 1} \eta_{jk}} \label{eqn:norm},
\end{align}
which is equivalent to the classical multinomial logistic regression model \citep[e.g.,][]{agresti_categorical_2002}. A graphical representation of the hierarchical model as a directed acyclic graph is available in the Appendix (Figure~\ref{fig:DAG}).

The hierarchical model linking satellite imagery, dominant vegetation, and landscape characteristics, offers several useful analytical outcomes. Estimation of the coefficients $\gamma_k$ provides insights into the seasonally varying reflectance properties of each species, which can highlight key differences among species and potentially reveal long-term changes to the seasonal patterns of green-up and senescence for vegetation. Estimation 
of $\beta_k$ can reveal how each species responds to variation in features like elevation and slope. Finally, estimation of $z_j$ yields probabilistic predictions of dominant vegetation for sites without ground truth observations. 

Data were simulated from the hierarchical model over a regular spatial 24 by 24 grid using 2 synthetic predictors ($P = 3$ including intercept) and model parameters drawn from priors closely aligned with beliefs about parameters in the context of the motivating application. Priors were were weakly informative in general, with one exception being strong positive dependence among basis coefficients corresponding to the same basis function for different categories ($\mathrm{Cov}(\gamma_{kl}, \gamma_{kl^*}) = 0.9, l \neq l*$) consistent with the knowledge that reflectance surfaces will be very similar across the $K = 3$ species under study. $L = 64$ basis functions were used to allow for sufficient flexibility in the shape of each expected reflectance surface. 

The 576 pixels in the simulation were partitioned randomly into 15 sets with a partition size of 36 sites for stage 1, 20 additional sites for stage 2, and 40 additional sites for stages 3--15.
Recursive Bayesian inference was implemented by first generating $M = 10,000$ samples from stage 1 with conventional MCMC methods, then applying SPP-RB for stages 2--15. Samples from the stage 1 transient posterior were generated using automated factor slice samplers \citep{neal2003slice} as implemented through the R package NIMBLE \cite{NIMBLEJCGS, NIMBLE}.

SPP-RB was carried out with $\lambda = 0$, as well as with a marginal KDE approach using Silverman's rule-of-thumb \citep{silverman_density_2018} as implemented by the function \verb=bw.nrw0()= in the R base package \verb=stats= \citep{R_2025}. Blocking was used to improve sampling efficiency, with blocking structures chosen through trial and error. All $P (K - 1) = 6$ regression coefficients, $\boldsymbol\beta$, were grouped into a single block. Reflectance coefficients, $\boldsymbol\gamma$, were grouped by associated basis function yielding $L = 64$ blocks of size $K = 3$ (i.e., $\{\gamma_{lk}: k = 1, \dots, K\}$). The variance parameter, $\sigma^2$, was assigned to its own block.

Compared to the logistic regression model in Section~\ref{sec:logistic}, the hierarchical model is more realistic setting in which to investigate the practical performance of the proposed SPP-RB approach. One fundamental distinction is the dimensionality of the parameter space. The hierarchical model includes $P(K-1) + LK + 1 = 100$ parameters compared to a maximum of 18 considered for logistic regression. In addition, the hierarchical model incorporates a complex posterior dependence structure among $\boldsymbol\beta$ and $\boldsymbol\gamma$, which could challenge the efficiency of the stage-wise MH algorithms, even with blocking. The hierarchical model was specially designed for the motivating application, and thus the performance of the SPP-RB approach with simulated data is expected to be a reliable indicator of its performance in the real application where an all-at-once reference posterior is computationally intractable.


\begin{figure}[h]
\includegraphics[width=\textwidth]{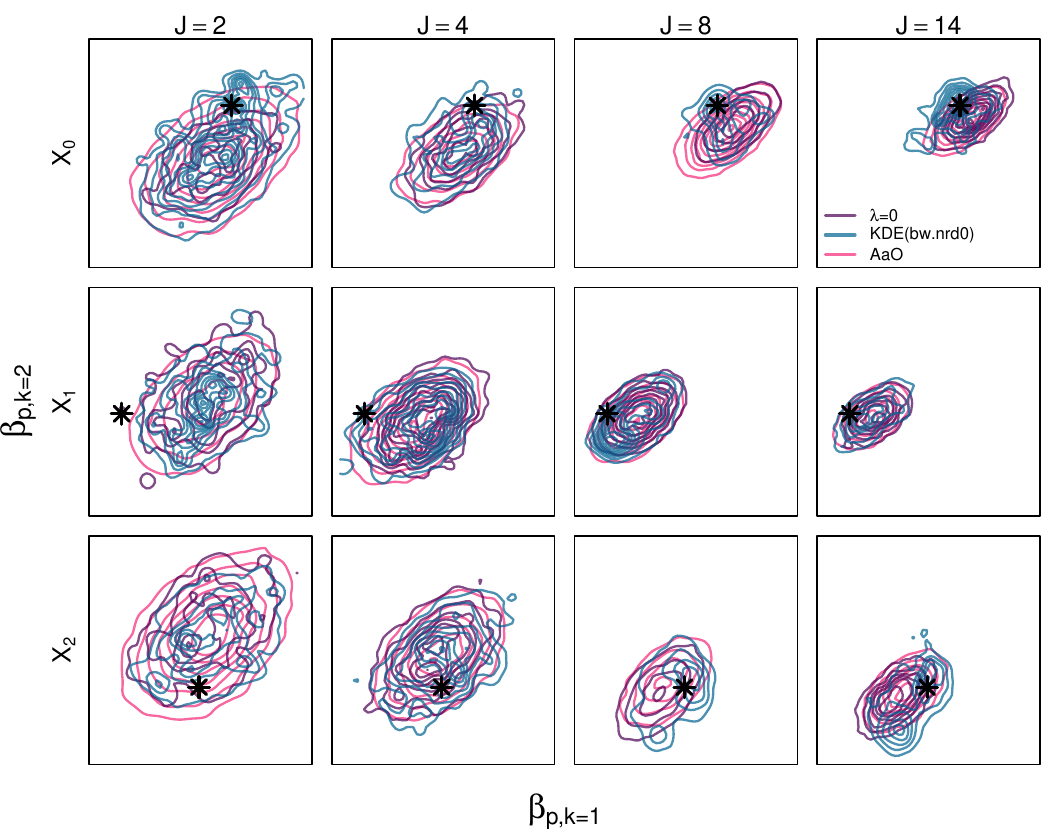}
\caption{Contour plots for transient posterior distributions. Each plot shows the bivariate distribution of two regression coefficients. Each pair are associated with one of $P = 3$ predictors, and correspond to the effects for categories $k = 1$ and $k = 2$ (with $k=3$ the implied baseline category). Colors correspond to posteriors obtained using SPP-RB with $\lambda = 0$ (purple), a marginal KDE approach (blue), and an all-at-once reference analysis (pink).}
\label{fig:beta_bivariate}
\end{figure}

Figure~\ref{fig:beta_bivariate} shows summaries of some transient posterior densities as bivariate contour plots. Within each column, a single plot shows the bivariate distribution of the relative covariate effects for species 1 and 2 relative to species 3 for a particular predictor (i.e., $\beta_{p,k=1}$ and $\beta{p,k=2}$ for $p \in 1, \dots, P=3$). Columns correspond to the stage associated with the transient posterior. SPP-RB with $\lambda = 0$ results in contours (purple) that are visually similar to those resulting from an all-at-once approach (pink). A marginal KDE approach yielded contours (blue) that are more visually distinct from the all-at-once baseline, suggesting that a multivariate normal approximation of transient posteriors might outperform marginal KDEs.

\begin{figure}[h]
\includegraphics[width=\textwidth]{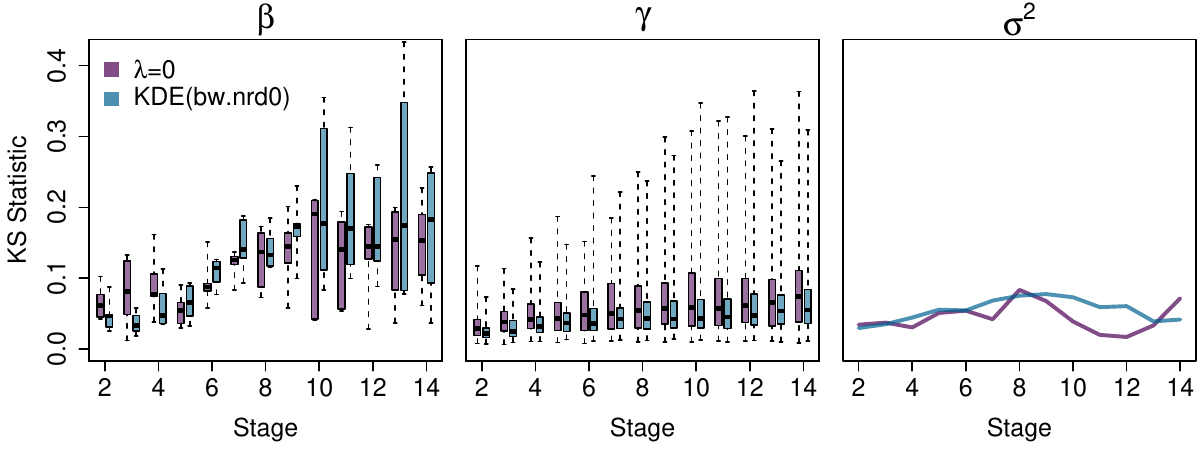}
\caption{Boxplots summarizing the discrepency between SPP-RB implementations and the all-at-once approach. Each boxplot represents the distribution of marginal Kolmogorov--Smirnov test statistics for each parameter in the model, with larger values indicating less agreement between the marginal SPP-RB and all-at-once posteriors. Sub-plots correspond to groups of parameters associated with landscape effects ($\boldsymbol\beta$), spatial random effects ($\boldsymbol\phi$), refletance surfaces ($\boldsymbol\gamma$), and reflectance observational error variance ($\sigma^2$).}
\label{fig:KS_SML_grouped}
\end{figure}

Figure~\ref{fig:KS_SML_grouped} shows boxplots summarizing the discrepancy between marginal SPP-RB transient posterior distributions and the corresponding transient posteriors using a direct all-at-once approach as measured by marginal KS statistics for each parameters. Each box represents $P(K-1) = 6$ parameters for the regression coefficients, $\boldsymbol\beta$, and $LK=192$ reflectance the surface coefficients, $\boldsymbol\gamma$. Figure~\ref{fig:KS_SML_grouped} suggests that SPP-RB with $\lambda = 0$ may slightly outperform an approach using marginal KDE for each parameter in the case of regression coefficients, $\boldsymbol\beta$, but the reverse may be true for $\boldsymbol\gamma$. 


Diagnostic plots for the case of $\lambda = 0$ analogous to Figure~\ref{fig:logistic_diagnostics} based on two different partition sequences suggest the SPP-RB method produces samples that generally align with the true ultimate posterior (Figures~\ref{fig:diag_sdm_beta} and \ref{fig:diag_sdm_gamma} in the Appendix).

\section{Discussion}\label{discussion}

SPP-RB provides a novel approach for implementing recursive Bayesian inference to obtain an approximate sample from the target posterior distribution, $[\boldsymbol\theta|\boldsymbol{y}]$, when an all-at-once approach is infeasible. By partitioning a massive data set, like ones often associated with satellite imagery, into batches, it is possible to break up a very large model fitting problem into a sequence of smaller, more manageable model fitting problems. The SPP-RB approach may be particularly valuable when data sets are large enough to overwhelm computating hardware's capacity for random access memory, as is the case for the motivating application. In addition, SPP-RB, like all recursive Bayesian approaches, provides a natural tool for the analysis of online data, allowing for updated posteriors without needing to revisit previously analyzed data. The primary contribution that SPP-RB makes compared to existing alternatives is a strategy for avoiding the accumulation of MC error that arises through particle depletion as the number of stages and unknown parameters grows. SPP-RB was shown to perform well in simulation for both simple settings such as logistic regression and more realistic hierarchical models such as the species distribution model developed for the motivating application.

One broad conclusion suggested by the simulation study is that SPP-RB works well in many settings for the case of $\lambda=0$, which corresponds to the assumption that all transient posteriors can be accurately represented using multivariate normal distributions. Such an approximation is supported by the result of the Benstein-von Mises theorem, which states that, under some conditions, posterior distribution converge asympotically to multivariate normal distributions. Although this empirical finding appears to obviate the need for more flexible distributions defined by $\lambda > 0$, the asymptotic rate of convergence to normality will generally be model and application sepecific.




\clearpage

\bibliographystyle{apalike}
\bibliography{recursiveBayes.bib}

\clearpage

\appendix

\input{appendix_content.tex}

\end{document}

%% file: appendix_content.tex
\section{Introduction}

\begin{figure}[h]
\includegraphics[width = \textwidth]{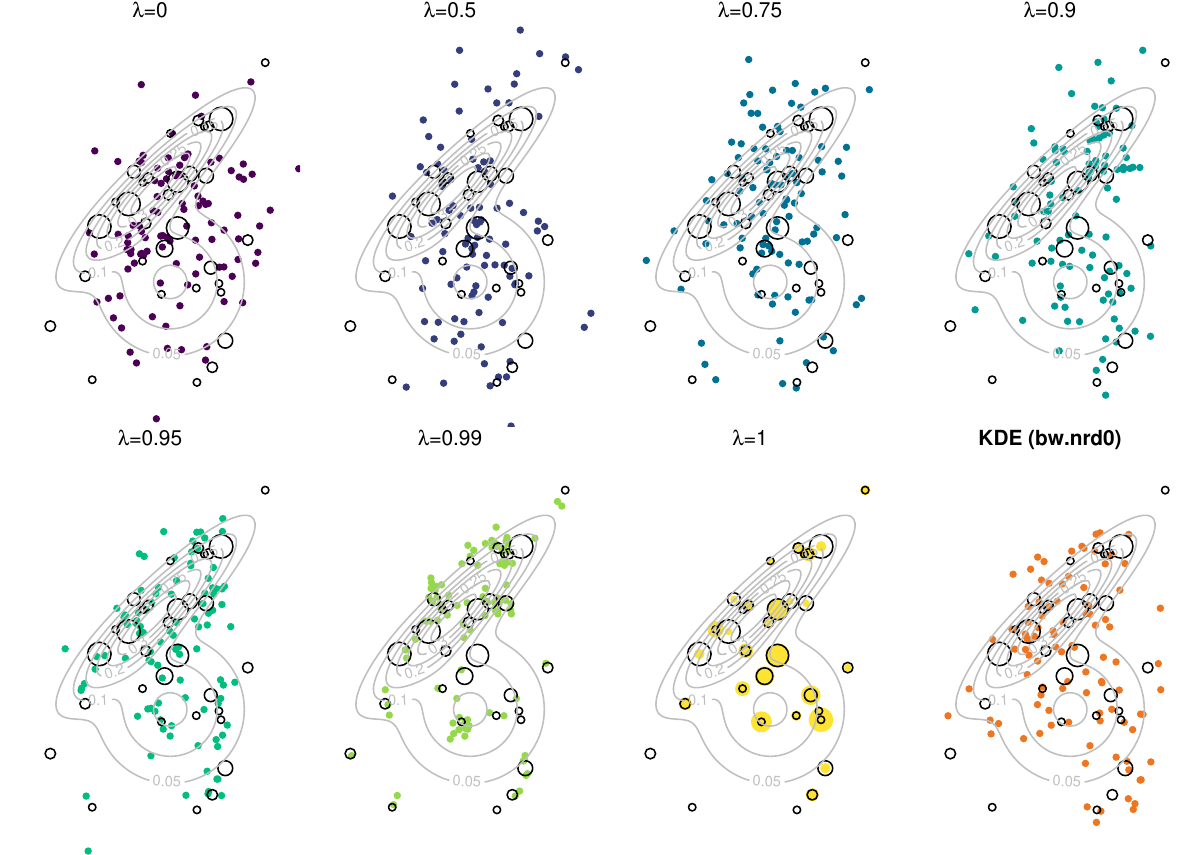}
\caption{Original (black, open) and perturbed (colors, filled) samples from a mixture of two bivariate Gaussian distributions for a variety of regularization levels ($\lambda$). Point sizes are scaled so that area increases linearly with multiplicity. Contour lines correspond to the original distribution (see Appendix for details). KDE figure shows perturbations using a diagonal bandwidth matrix.}
\label{fig:reg_KDE}
\end{figure}

Figure~\ref{fig:reg_KDE} shows examples of 100 perturbations of an artificially depleted sample originally arising from the mixture of two Gaussian distributions. Several values of $\lambda$ between 0 (jointly Gaussian) and 1 (multinomial resampling) are shown, along with a KDE based on a diagonal bandwidth matrix. Note the smaller number of unique points for multinomial resampling ($\lambda = 1$).

\subsection{Figure~\ref{fig:reg_KDE} details}
Original samples were drawn according to the following two-step procedure that induces multiplicity of values.
\begin{enumerate}
\item Draw $50$ realizations from each of two mixture distributions, 
$\mathrm{N}\!\left(\begin{pmatrix}0.25 \\ 0 \end{pmatrix}, \begin{pmatrix}1 & 0 \\ 0 & 1  \end{pmatrix}\right)$ and 
$\mathrm{N}\!\left(\begin{pmatrix}0 \\ 2 \end{pmatrix}, \begin{pmatrix}1 & 0.9 \\ 0.9 & 1 \end{pmatrix}\right)$, 
and combine to form preliminary sample denoted $\{\boldsymbol\theta^{i(1)}\}_{i = 1}^{M=100}$.
\item For $j$ in $2, \dots, 5$, sample uniformly from $\{\boldsymbol\theta^{i(j-1)}\}_{i = 1}^M$ with replacement. Denote the new sample $\{\boldsymbol\theta^{i(j)}\}_{i = 1}^M$ and repeat.
\end{enumerate}
The procedure is equivalent to recursively refreshing 4 times according to $\lambda = 1$. The final number of unique realizations was 32.

Samples $\{\boldsymbol\theta_{i(5)}\}_{i = 1}^M$ were subsequently refreshed for $\lambda \in \{0, 0.5, 0.75, 0.9, 0.95, 0.99, 1\}$ and using a naive KDE approach with marginal bandwidths for each variable selected using the function \verb=bw.nrd0()= with input $\{\boldsymbol\theta_{i(5)}\}_{i = 1}^M$.

\section{Methods}

\subsection{Asymptotic equality of $\mathrm{Var}(\boldsymbol\theta^*)$ and $\mathrm{Var}(\boldsymbol\theta)$}
The following shows that when $\boldsymbol\theta^*|\boldsymbol{\tilde\theta}, \{\boldsymbol\theta^i\}_{i = 1}^M$ is defined according to \eqref{eqn:reg_KDE}, $\lim_{M \rightarrow \infty}\mathrm{Var}(\boldsymbol\theta^*) = \mathrm{Var}(\boldsymbol\theta)$. That is, provided there is a large enough sample size, $M$, at each stage, the marginal variances of the transient posterior and perturbed sample are approximately equal.

\textbf{Note}: For $\boldsymbol{\tilde\theta}|\{\boldsymbol\theta\}^M \sim \mathrm{Cat}\left(\{\boldsymbol\theta\}^M\right)$, and each $\boldsymbol\theta^i$ having shared marginal distribution $[\boldsymbol\theta]$, the marginal distribution of $\boldsymbol{\tilde\theta}$ is also $[\boldsymbol\theta]$, and thus $\mathrm{Var}(\boldsymbol{\tilde\theta}) = \mathrm{Var}(\boldsymbol\theta)$, regardless of the value of $M$ or dependence among elements of $\{\boldsymbol\theta\}^M$.

The variance of $\boldsymbol\theta^*$ can be derived using the law of total variation.
\begin{align*}
\mathrm{Var}(\boldsymbol\theta^*) &= 
  \mathrm{E}\left(\mathrm{Var}(\boldsymbol\theta^* | 
                                 \boldsymbol{\tilde\theta}, \{\boldsymbol\theta\}^M)\right) + 
  \mathrm{Var}\left(\mathrm{E}(\boldsymbol\theta^* | 
                                 \boldsymbol{\tilde\theta}, \{\boldsymbol\theta\}^M)\right) \\
&= \mathrm{E}(\mathbf{S}_\theta - \boldsymbol\Lambda \mathbf{S}_\theta \boldsymbol\Lambda') + 
    \mathrm{Var}(\boldsymbol\Lambda\boldsymbol{\tilde\theta} + 
      (\mathbf{I} - \boldsymbol\Lambda)\boldsymbol{\bar\theta}) \\
  &= \mathrm{Var}(\boldsymbol\theta) - 
    \boldsymbol\Lambda\mathrm{Var}(\boldsymbol\theta)\boldsymbol\Lambda' + 
                \left(\boldsymbol\Lambda + M^{-1}(\mathbf{I} - \boldsymbol\Lambda)\right)
              \mathrm{Var}(\boldsymbol{\tilde\theta})
              \left(\boldsymbol\Lambda + M^{-1}(\mathbf{I} - \boldsymbol\Lambda)\right)' + \\
    & \qquad M^{-2}(\mathbf{I} - \boldsymbol\Lambda)\mathrm{Var}\left(
      \sum_{\boldsymbol\theta^i \neq \boldsymbol{\tilde\theta}} \boldsymbol\theta^i\right)
    (\mathbf{I} - \boldsymbol\Lambda)'\\
              &= \mathrm{Var}(\boldsymbol\theta) - 
                \boldsymbol\Lambda\mathrm{Var}(\boldsymbol\theta)\boldsymbol\Lambda' +
    \boldsymbol\Lambda\mathrm{Var}(\boldsymbol\theta)\boldsymbol\Lambda' + \\
              &\qquad M^{-1}\boldsymbol\Lambda\mathrm{Var}(\boldsymbol\theta)(\mathbf{I} - \boldsymbol\Lambda)' + 
    M^{-1}(\mathbf{I} - \boldsymbol\Lambda)\mathrm{Var}(\boldsymbol\theta)\boldsymbol\Lambda' +\\
              &\qquad (M - 1)M^{-2}(\mathbf{I} - \boldsymbol\Lambda)\mathrm{Var}
              (\boldsymbol\theta)(\mathbf{I} - \boldsymbol\Lambda)' + \\
    &\qquad M^{-2}(\mathbf{I} - \boldsymbol\Lambda)
      \sum_{\boldsymbol\theta^i \neq \boldsymbol\theta^j \neq \boldsymbol{\tilde\theta}}
        \mathrm{Cov}(\boldsymbol\theta^i, \boldsymbol\theta^j)
      (\mathbf{I} - \boldsymbol\Lambda)'
              \end{align*}
              Taking the limit of both sides as $M \rightarrow \infty$, several terms in the sum clearly have limits of 0, leaving
              \begin{align}
              \lim_{M\rightarrow\infty}\mathrm{Var}(\boldsymbol\theta^*) = 
                \mathrm{Var}(\boldsymbol\theta) + \lim_{M\rightarrow\infty}
              M^{-2}(\mathbf{I} - \boldsymbol\Lambda)
              \sum_{\boldsymbol\theta^i \neq \boldsymbol\theta^j \neq \boldsymbol{\tilde\theta}}
              \mathrm{Cov}(\boldsymbol\theta^i, \boldsymbol\theta^j)
              (\mathbf{I} - \boldsymbol\Lambda)'.
\end{align}
As long as the dependence in $\{\boldsymbol\theta\}^M$ is controlled entrywise such that 
\begin{align}
\left[\sum_{\boldsymbol\theta^i \neq \boldsymbol\theta^j}
  \mathrm{Cov}(\boldsymbol\theta^i, \boldsymbol\theta^j)\right]_{l,m} < \mathcal{O}(M^2)
  \;\; \forall \;\; l, m = 1, \dots, M,
\end{align}
or equivalently that the individual entries in $\sum_{i = 1}^M \mathrm{Cov}(\boldsymbol\theta^i, \boldsymbol\theta^{i + h})$ are $< \mathcal{O}(M)$, the RHS will reduce to $\mathrm{Var}(\boldsymbol\theta)$. This requirment is satisfied for many commonly studied covariance functions (e.g., autoregressive processes, Mat\'{e}rn covariance family).


%

\clearpage

\section{Simulation study}

\subsection{Logistic regression}


\begin{figure}[h]
\includegraphics[width=\textwidth]{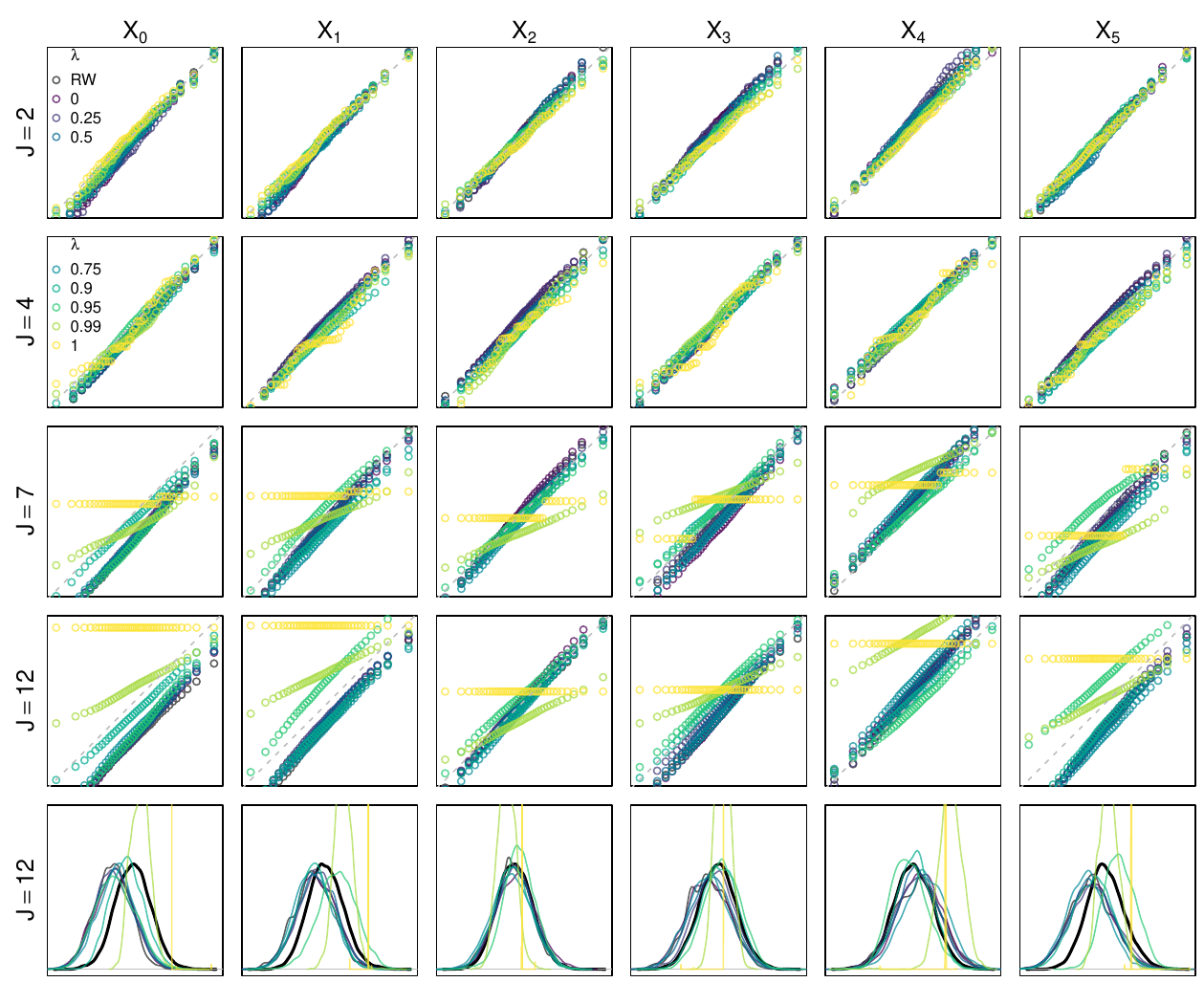}
\caption{Quantile-quantile plots comparing each SPP-RB-derived transient posterior to that obtained via the all-at-once approach. Bottom row shows marginal KDEs of ultimate posteriors.}
\label{fig:qq_spprb}
\end{figure}

\begin{figure}[h]
\includegraphics[width=\textwidth]{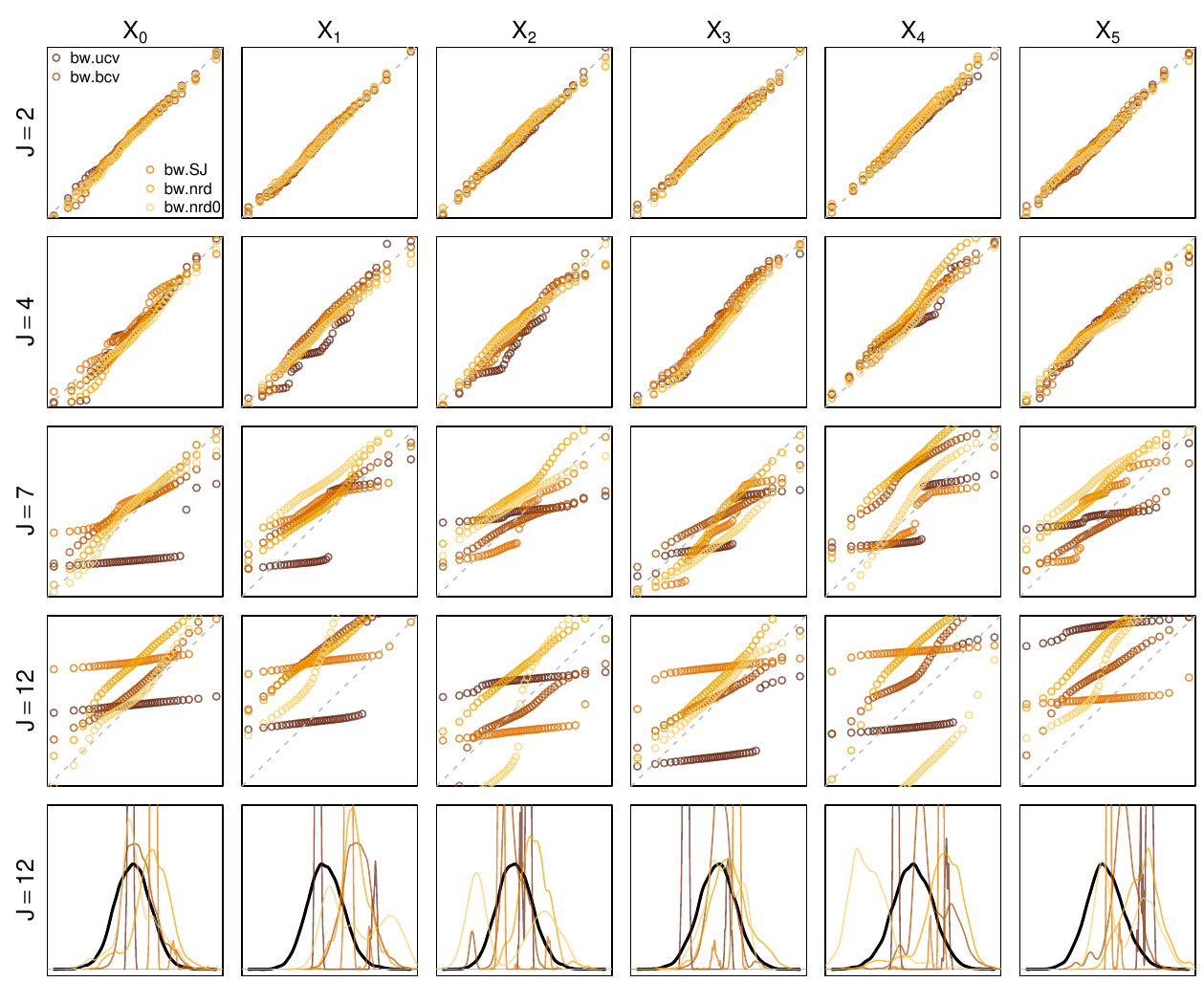}
\caption{Quantile-quantile plots comparing each SPP-RB-derived transient posterior to that obtained via the all-at-once approach. Bottom row shows marginal KDEs of ultimate posteriors.}
\label{fig:qq_kde}
\end{figure}

%

\begin{figure}[h]
\includegraphics[width=\textwidth]{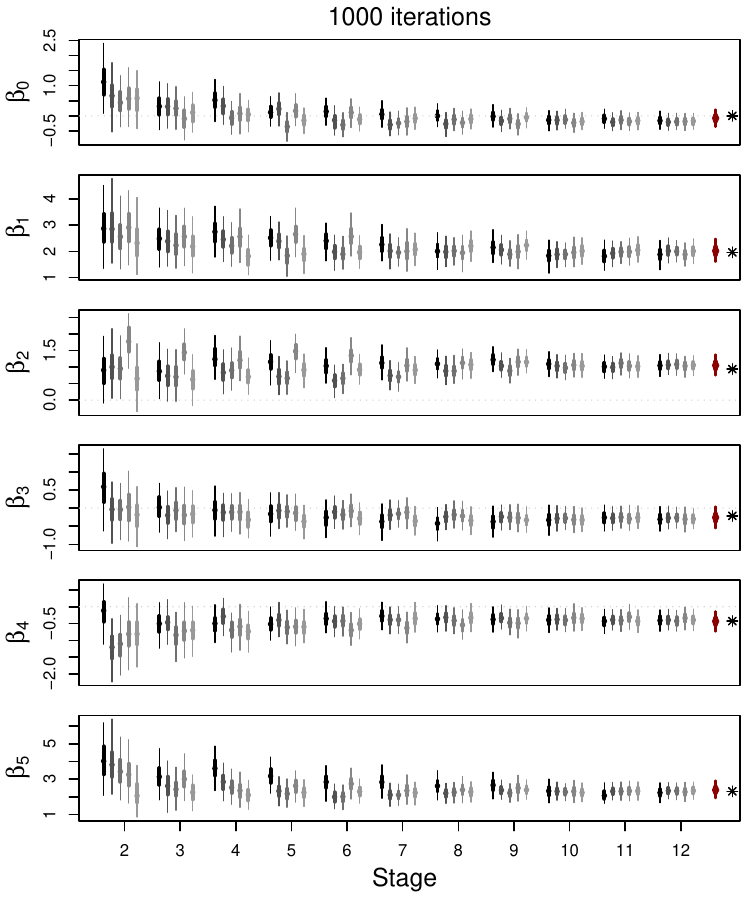}
\caption{See Figure~\ref{fig:logistic_diagnostics}.}
\label{fig:diag_logistic}
\end{figure}

\clearpage

\subsection{Species distribution model}

\begin{figure}[h]
\centering
\input{SML_dag_no_phi.tex}
\caption{Directed acyclic graph representation of hierarchical species distribution model. Circular nodes represent stochastic elements and rectangular nodes represent fixed, known elements. Solid lines represent relationships defined by a conditional probability distribution, and dotted lines represent deterministic relationships. Shading emphasizes hierarchy in model structure, with cover type acting as a link between landscape characteristics and reflectances.}
\label{fig:DAG}
\end{figure}
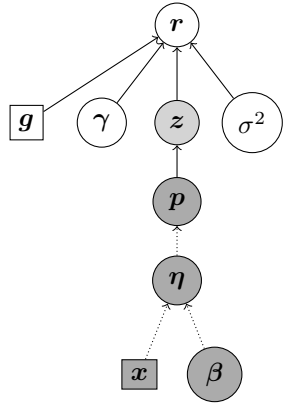

\begin{figure}[h]
\includegraphics[width=\textwidth]{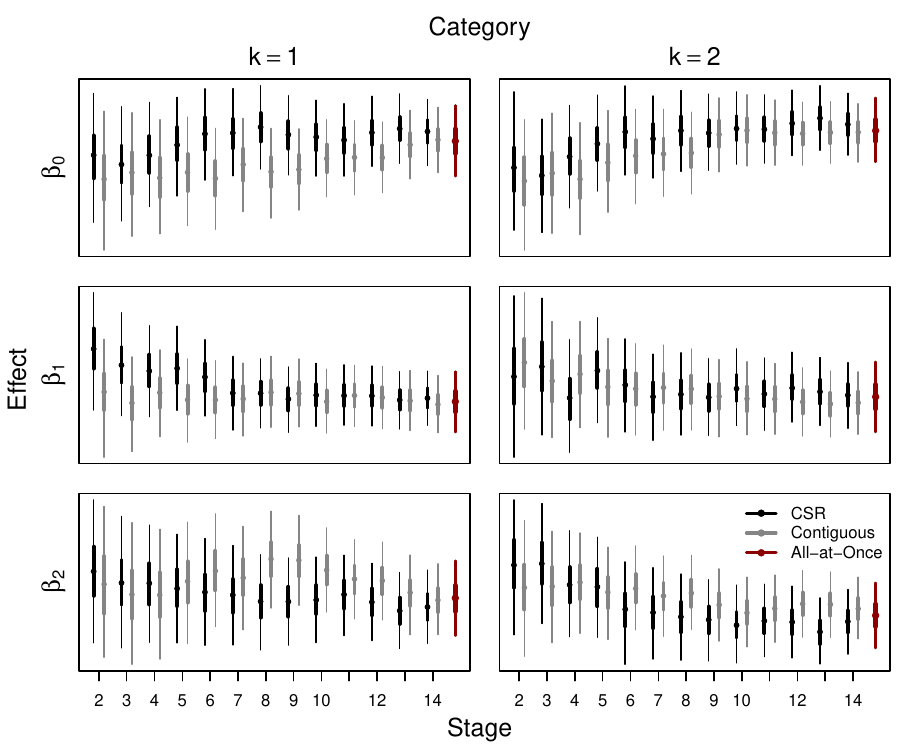}
\caption{Boxplots summarizing the marginal transient posterior densities for $\boldsymbol\beta$ in the species distribution model. Black and gray represent two distinct partitions underlying two implementations of SPP-RB.}
\label{fig:diag_sdm_beta}
\end{figure}

\begin{figure}[h]
\includegraphics[width=\textwidth]{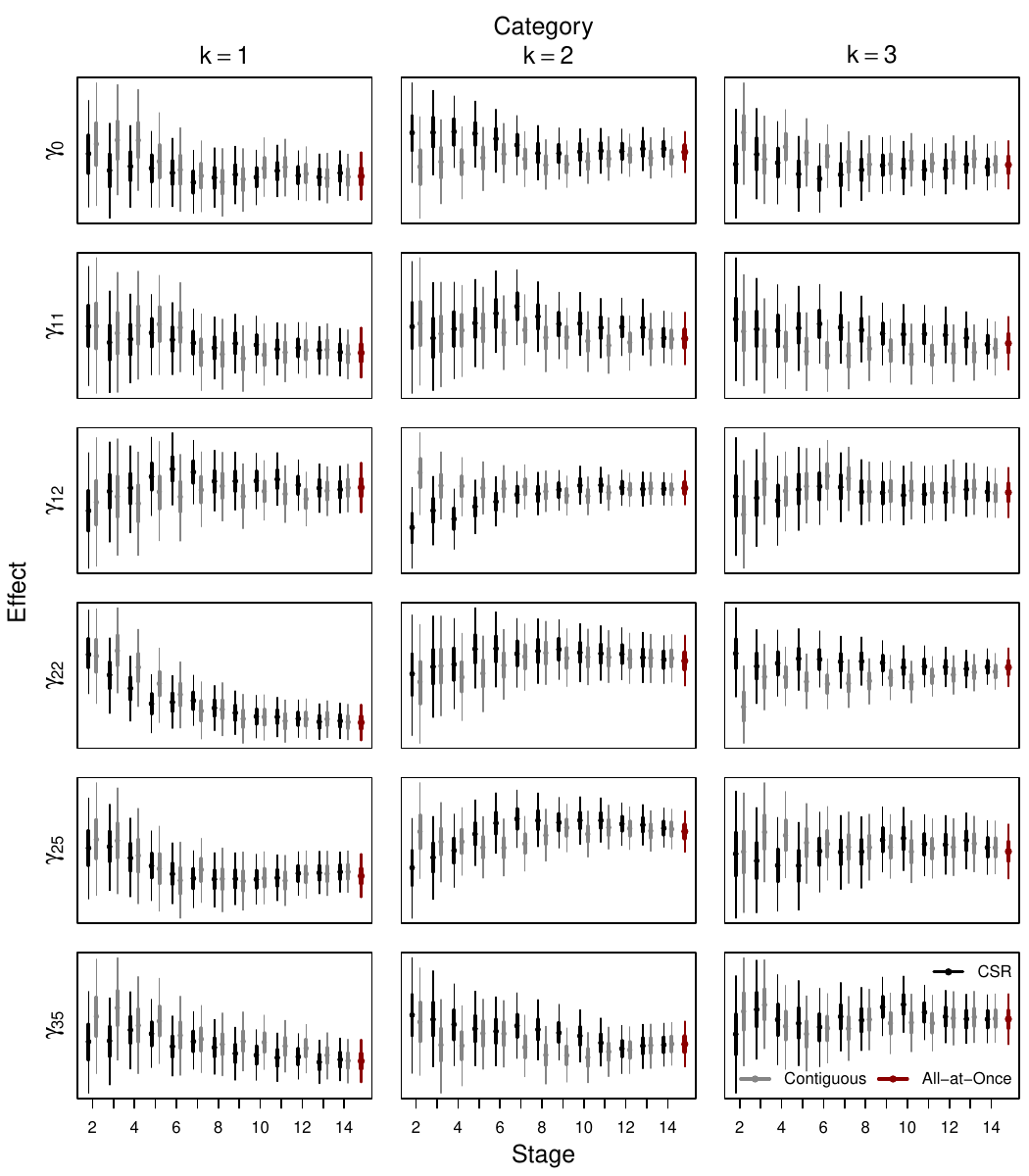}
\caption{Boxplots summarizing the marginal transient posterior densities for ($\boldsymbol\gamma$) in the species distribution model. Black and gray represent two distinct partitions underlying two implementations of SPP-RB.}
\label{fig:diag_sdm_gamma}
\end{figure}

%% file: SML_dag_no_phi.tex
\begin{tikzpicture}
    \node[shape=circle,draw=black] (r) at (2.5,0) {$\boldsymbol{r}$};
    
    \node[shape=rectangle,draw=black] (g) at (0.5,-1.3) {$\boldsymbol{g}$};
    \node[shape=circle,draw=black] (gamma) at (1.5,-1.3) {$\boldsymbol\gamma$};
    \node[shape=circle,draw=black,fill=black!17] (z) at (2.5,-1.3) {$\boldsymbol{z}$};
    \node[shape=circle,draw=black] (sigma) at (3.5,-1.3) {$\sigma^2$};
  
    \node[shape=circle,draw=black,fill=black!33] (prob) at (2.5,-2.35) {$\boldsymbol{p}$};
    
    \node[shape=circle,draw=black,fill=black!33] (eta) at (2.5,-3.4) {$\boldsymbol\eta$};

    \node[shape=rectangle,draw=black,fill=black!33] (x) at (2,-4.65) {$\boldsymbol{x}$};
    \node[shape=circle,draw=black,fill=black!33] (beta) at (3,-4.65) {$\boldsymbol\beta$};

    \path [->] (z) edge node[left] {} (r);
    \path [->] (g) edge node[left] {} (r);
    \path [->] (gamma) edge node[right] {} (r);   
    
    \path [->] (sigma) edge node[right] {} (r);
    
    \path [->] (prob) edge node[right] {} (z);
    
    \path [->, densely dotted] (eta) edge node[right] {} (prob);

    \path [->, densely dotted] (x) edge node[left] {} (eta);
    \path [->, densely dotted] (beta) edge node[right] {} (eta);
\end{tikzpicture}